# PATIENT SPECIFIC NUMERICAL SIMULATION OF FLOW IN THE HUMAN UPPER AIRWAYS FOR ASSESSING THE EFFECT OF NASAL SURGERY

Maria R. JORDAL[1*], Sverre G. JOHNSEN[2], Sigrid K. DAHL[2], Bernhard MÜLLER[1]

[1] NTNU Department of Energy and Process Engineering, 7491 Trondheim, NORWAY
[2] SINTEF Materials and Chemistry, 7465 Trondheim, NORWAY

* E-mail: mariarjordal@gmail.com

## ABSTRACT
The study is looking into the potential of using computational fluid dynamics (CFD) as a tool for predicting the outcome of surgery for alleviation of obstructive sleep apnea syndrome (OSAS). From pre- and post-operative computed tomography (CT) of an OSAS patient, the pre- and post-operative geometries of the patient's upper airways were generated. CFD simulations of laminar flow in the patient's upper airway show that after nasal surgery the mass flow is more evenly distributed between the two nasal cavities and the pressure drop over the nasal cavity has increased. The pressure change is contrary to clinical measurements that the CFD results have been compared with, and this is most likely related to the earlier steps of modelling – CT acquisition and geometry retrieval.

## NOMENCLATURE
*Greek Symbols*
$\rho$  Mass density, [kg/m$^3$].
$\mu$  Dynamic viscosity, [Pa s].

*Latin Symbols*
A  Cross sectional area [m$^2$].
$D_H$  Hydraulic diameter [m].
p  Pressure, [Pa].
P  Perimeter [m].
Q  Volumetric flow rate, [ml/s].
R  Resistance, [Pa s/ml].
$U_{avg}$  Average velocity [m/s]
**V**  Velocity vector, [m/s].

*Abbreviations*
AHI  Apnea-hypopnea index
CFD  Computational fluid dynamics
CT  Computed tomography
HU  Hounsfield units
OSAS  Obstructive sleep apnea syndrome
PNIF  Peak nasal inspirational flow
RANS  Reynold averaged Navier Stokes
RMM  Rhinomanometry
RRM  Rhinoresistometry

## INTRODUCTION
Obstructive sleep apnea syndrome (OSAS) is a disorder characterized by repeated collapses of the upper airways, preventing air from flowing freely during sleep, causing apneas (pauses in breath) and hypopneas (shallow breathing). The severity of sleep apnea is indicated by the number of apnea/hypopnea events per hour during sleep, which defines the apnea-hypopnea index (AHI), where <5 is considered normal and >30 severe. The most prevalent symptoms are daytime sleepiness, unrefreshing sleep and snoring, but OSAS has also been shown to increase the chance of cardiovascular diseases (AASM, 1999).

Several surgical and non-surgical treatment options exist for alleviation of OSAS, but it is difficult to predict the outcomes of the treatments. As the success rates of the treatments are highly varying from patient to patient, a tool for predicting their outcome is needed. CFD may aid as such a tool, and may provide a non-invasive and cost-efficient guidance to medical personnel on what surgery procedure to choose.

### Outline
In the current paper we have simulated the flow in the upper airways of one OSAS patient before and after intranasal surgery. The work is based on the treatments for OSAS at St. Olav University Hospital in Trondheim, Norway. Here, intranasal surgery is being performed on patients with OSAS. Only one third of the patients experience improvement in OSAS after surgery. It is not known why there is such a low success rate after surgery, and why some patients improve and others do not (Moxness and Nordgård, 2014). By studying the geometry and flow patterns of the upper airways before and after surgery, the impact of intranasal surgery on the airflow in the upper airway might become clearer. The method for creating computational models from CT images follows in the next section. Selected results (pressure, velocity and nasal resistance) will be discussed and compared with measured results. This article is based on the M.Sc. thesis by Jordal (2016).

A schematic of the upper airways and definitions of the anatomical planes and directions can be seen in Fig. 1.

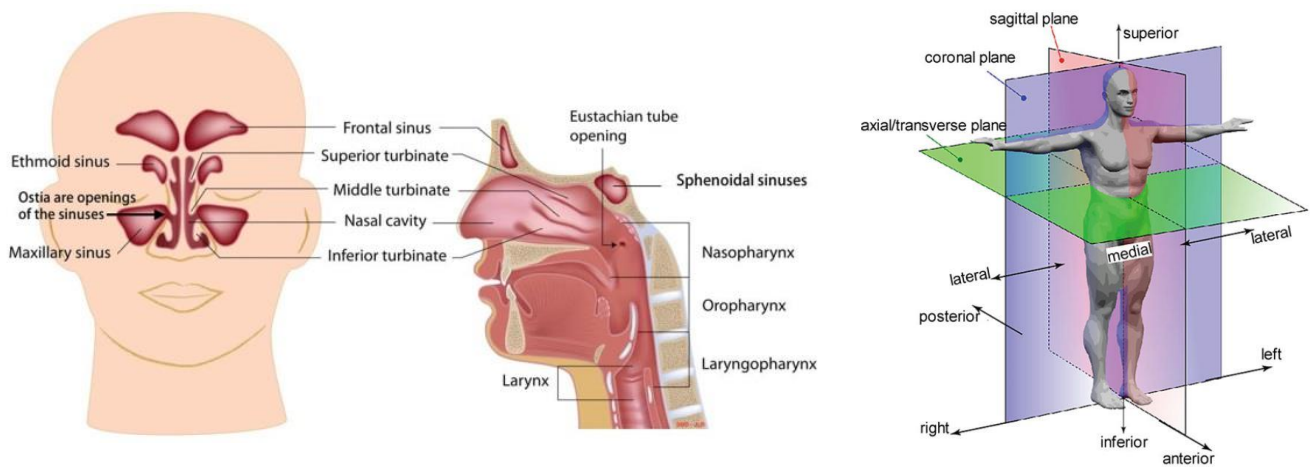

**Figure 1**: Schematic of the upper airways (Southlake sinus and snoring center, 2017) and of the anatomical planes and directions (Tu et al., 2013).

## METHOD

### Geometry Retrieval

*Data Acquisition*

The pre- and post-operative geometries were reconstructed from CT images provided by the Department of Radiology and Nuclear Medicine at St. Olav University Hospital, Trondheim. The CT was done with a Siemens Sensation 64 in the transverse plane. The pre-operative scan provided 342 slices with a slice thickness of 1.0 mm, and the post-operative scan provided a total of 423 slices with a slice thickness of 1.5 mm. All of the 2D CT images consisted of 512x512 pixels.

*Patient Data*

The patient chosen for this particular study is a man born in 1948 with a body mass index of 28. He underwent intranasal surgery at St. Olav Hospital in the fall of 2015 for alleviation of OSAS. The patient had a narrow nasal passage in his left nostril obstructing the airflow, and had surgery to increase the volume of this passage. A result of this intranasal surgery was a reduction in AHI from 23 to 5.7. As AHI<5 is considered normal, this indicates that the patient is almost alleviated of OSAS.

*Segmentation Procedure and Editing of Geometry*

The segmentation of the upper airways was done using ITK-SNAP 3.4.0 (Yushkevich, 2006). The automatic segmentation was performed using the Active Contour Method with thresholding. Air defines the lower limit of the Hounsfield Unit (HU)-scale at -1024HU, but there is no standard as to what the upper limit should be. Upper HU-values such as -300 (Ito et al., 2011), -400 (De Backer et al., 2007), -460 to -470 (Nakano et al., 2013) and -587 (Weissheimer et al., 2012) have been used for automatic segmentation in previous works. Although there is a big range of the upper limit, all of the above mentioned reported good results with these settings. For this segmentation, -300 as the upper HU-value has been chosen based upon trial and error (Jordal, 2015).

In addition, manual segmentation was necessary in order to capture the geometry. For this, the paint brush mode was used on the slices in all planes (coronal, axial and sagittal). For simplicity of the model, the paranasal sinuses were excluded from the model. The entire segmentation process was done in cooperation with an ear-nose-throat surgeon and a radiologist to make sure the model was anatomically correct.

The segmented volume was extracted from ITK-SNAP as a triangulated surface mesh. Netfabb basic (Nettfabb basic v.7.3) was used to analyze and check the quality of the mesh, and MeshLab (Cignoni et al., 2008) for further post-processing. In MeshLab, the mesh was reduced using the built-in function Quadratic Edge Collapse Decimation with topology preservation and a target number of faces of 100 000. This reduces the size of the mesh and the size of the file, which all reduces the time on editing the geometry in the steps that follows. Finally, the mesh was smoothed using the Laplacian Smooth Filter with default settings to avoid any artefacts from digitalization.

Since the patient had been positioned differently on the pre- and post-operative CT the pharynx appeared rather different for the two models (Fig. 2). The bend in the neck post-operatively is a result of a head-rest that was used during CT. The result of this is that the angle between the nasal cavity and the pharynx is larger post-operatively, in addition to some changes in the pharynx and larynx as the walls are elastic.

However, the only difference between the two models should between geometry of the nasal cavities. To make sure the only difference between the two models was the surgery, and avoid any effects that may be caused from the different positioning during CT, the post-operative nasal cavity was combined with the pre-operative pharynx and larynx. In order to do this, the models were first aligned in MeshLab, converted from surface mesh (stl file) to a solid body (stp file) in ANSYS Spaceclaim, and then combined in ANSYS DesignModeler. The two models had different circumferences, and in order to join the two parts together without creating a stair-step, a small volume (length of 3 mm.) between the two parts was created. In addition, the outlet was extended by cutting the model at the larynx and extending it in the flow direction using ANSYS DesignModeler in order to avoid reverse flow and to smooth the air flow at the outlet (Fig. 3).

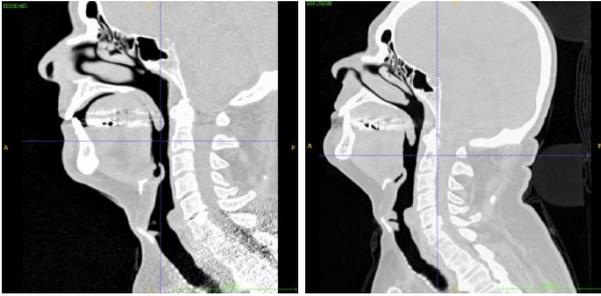

**Figure 2**: Pre- and post-operative CT images of the patient showing the difference in head positioning, respectively. Sagittal view from the left side.

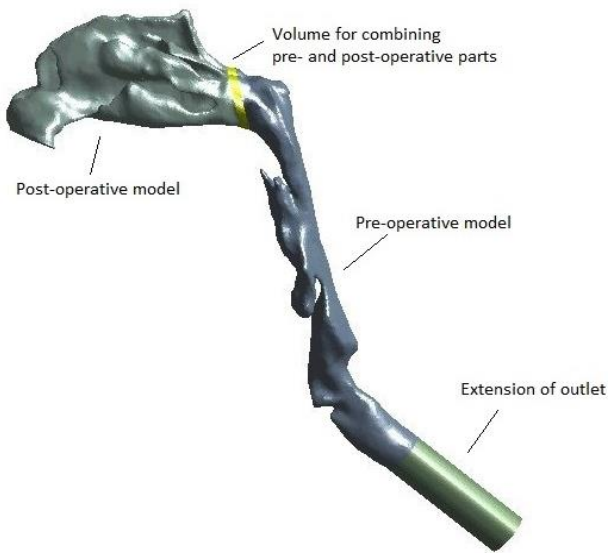

**Figure 3**: The final post-operative geometry showing the different parts that were combined into one single geometry. Viewed from the left side.

**Grid Generation**

A grid convergence test was carried out on tetrahedral and polyhedral grids made in Ansys Meshing (ANSYS inc, v.16.2). Coarse, medium and fine (mostly) tetrahedral grids were made in Ansys Meshing by choosing "no set cell type" as the cell type in Ansys Meshing. The same grids were then converted to polyhedral cells in Ansys Fluent. In addition to these six grids, a coarse grid with inflation layers was included in the test. The computational time was about 30% lower on the polyhedral grids, and a medium type grid showed grid independence. Based on the grid convergence test, grids were made for both pre- and post-operative models. The post-operative grid was made with the medium settings which resulted in a grid with 19 783 513 nodes and 3 489 365 polyhedral cells. The pre-operative grid was then made to approximately match the number of cells, and consists of 17 023 087 nodes and 2 993 762 polyhedral cells.

**Numerical Simulation**

The incompressible Navier-Stokes equations were solved for the entire domain. They read:

$$\text{div } \mathbf{V} = 0 \quad (1)$$

$$\rho \frac{D\mathbf{V}}{Dt} = -\nabla p + \mu \nabla^2 \mathbf{V} \quad (2)$$

where $\mathbf{V}$, $p$, $\rho$, $\mu$ are the velocity vector, pressure, mass density and dynamic viscosity, respectively. The software ANSYS Fluent was used for numerical simulations. The flow simulated is modelled as incompressible, hence the pressure-based solver was chosen. This solver is also default in Fluent. The pressure based solver couples the velocity and pressure and for this SIMPLE (semi-implicit method for pressure-linked equations) was chosen. SIMPLE is default in ANSYS Fluent. For the spatial discretization, the following default settings were used; Gradient: Least square cell based, pressure: second order, momentum: second order upwind. For the transient formulation (when applied), the second order implicit was used. Inspirational flow was simulated by defining the nostrils as inlets with atmospheric pressure (0 Pa total pressure), and the end of larynx as the outlet with a uniform outflow velocity corresponding to 250ml/s. The no-slip condition was applied at the walls. The flow was simulated as laminar with $\rho = 1.225$ kg/m$^3$ and $\mu = 1.7894 \cdot 10^{-5}$ $Pas$.

Because of the large amount of grid cells, the simulations were done on a high performance computer available at NTNU. With 12 CPUs, this took about two days, but the solution was not fully stable. The flow was modelled as steady-state, but to reach a solution, the flow was solved as transient with a time step of 10e-6 seconds with a maximum of 20 iterations per time step. This went on until the solution converged with the scaled residuals in the order of e-09 to e-13.

**Clinical Measurements**

From St. Olav Hospital, data from rhinometric measurements, such as rhinoresistometry (RRM) and rhinomanometry (RMM) were available. RRM and RMM measure the resistance in the nose at different flow rates. The resistance, R, is defined as R = ΔP/Q where ΔP is the pressure difference from the nostrils to the posterior nose/beginning of nasopharynx, and Q is the volumetric flow rate. The resistances of the left and right nasal cavities are measured individually. The test procedure is to close one of the nostrils, placing a mask over the nose and mouth, and letting the patient breathe in and out at normal pace. All tests have been done both pre- and post-operatively, before and after decongestion of the nose. The tests were first taken when the nose was at its normal state. After this, the patient was given nasal spray, and waited 15 minutes before the tests were retaken. This was done to decongest the nose and eliminate the effect of mucosa

To compare the measured values, the resistance was calculated from the CFD-results. The volumetric flow rate was calculated from the mass flow rates of each of the nostrils, and the pressure drop was defined from the inlets (nostrils) to the posterior nose. The results were compared with the results from RMM and RRM at the same flow rates.

## RESULTS
### Geometry
The difference between the two final geometries is the nasal cavity, and a more detailed view of this can be seen in Fig. 4-6.

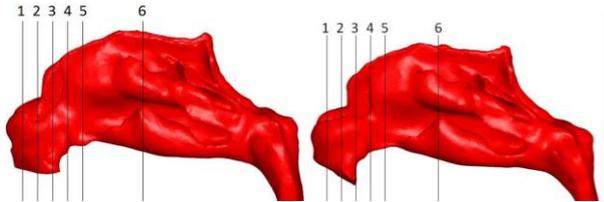

**Figure 4**: The nasal cavity pre(left)- and post(right)-operative viewed from the left side. Planes used for cross sections in Fig. 5-6 are marked.

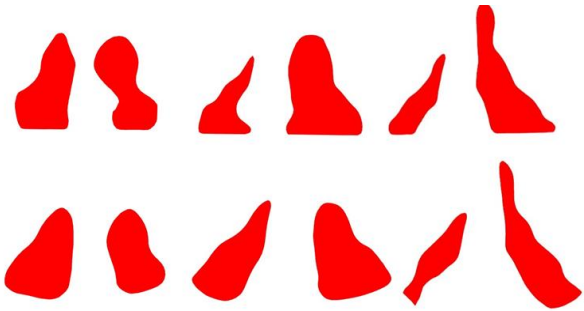

**Figure 5**: Cross sections at planes 1-3. Pre-operative model on top, and post-operative below.

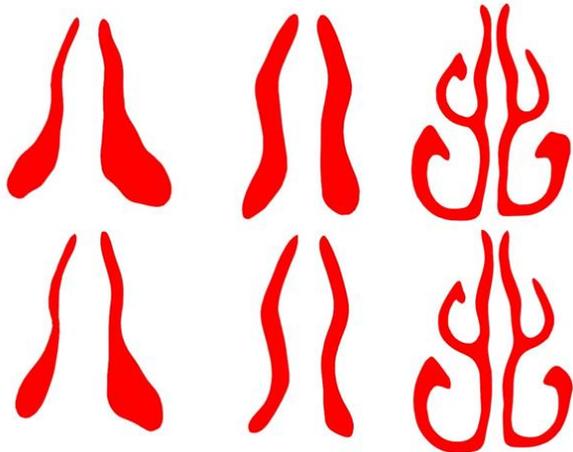

**Figure 6**: Cross sections at planes 4-6. Pre-operative model on top and post-operative model below.

The surgery was performed in order to open up the narrow airway on the anterior left nasal cavity, and straighten out the septum. As seen in Fig. 5-6, this volume has increased on both left and right side in the anterior nose. However, the volume appears to have decreased after surgery posterior in the nose. This will be discussed further. The inlets of the pre- and post-operative models are angled slightly different, but this is not affecting the air flow remarkably (Taylor et.al. 2010).

### Velocity
The velocity distribution in the upper airways both pre- and post-operatively showed lower velocities in the nasal cavity, and an increase in velocity as the cross sectional area becomes narrower in the pharynx. The velocity magnitudes across a sagittal cut plane can be seen in Fig.10. The plane is positioned in the middle of the pharynx and the larynx, and close to the septum on the left nasal cavity.

The highest velocities are found in the smallest cross sectional area, which is behind the epiglottis. This narrowing creates a pharyngeal jet. A large change in the angle between the pharynx and larynx creates swirling and recirculation in the larynx. The maximum velocities are almost identical pre- and post-operative at 7.783 and 7.827 m/s, respectively. As the mass flow is constant and identical in both cases, and the pre- and post-operative geometry is the same from the nasopharynx and below, the velocity is expected to be similar in these areas. Differences in the pre- and post-operative nasal cavities can be observed as the geometry has changed (Fig. 9).

It can be seen that the velocity magnitude has increased on the left side after surgery. The highest velocities are found in the inferior nasal cavity around the inferior turbinate and close to the septum. The lowest velocities are observed in the olfactory zone and at the edges. The findings correspond with the description of flow patterns in the literature (Schreck et al., 1993, Hahn et al., 1993, Keyhani et al., 1995). The flow is more evenly spread out in the right nasal cavity. The differences between the left and right nasal cavity and the pre- and post-operative nasal cavities are well illustrated by the velocity streamlines (Fig. 11).

In Fig. 11 it can be seen that the majority of the flow is in the inferior nasal cavity. After surgery, the velocities are higher in the left nasal cavity. This is a result of an increased volume in the left anterior nasal cavity allowing more air in. Before surgery 15% more of the flow went through the right nasal cavity than the left. After surgery, this difference is reduced to 8%. Even though the mass flow rate in the right nasal cavity is lower after surgery, the velocity has not decreased. However, the cross sectional area appear to have decreased. This will be discussed further below.

From the velocity, the Reynolds Number was calculated at the cross sections marked in Fig. 8 as follows:

$$Re = \frac{\rho U_{avg} D_H}{\mu} \quad (3)$$

where $D_H$ is the hydraulic diameter, $D_H = 4A/P$, where A is the cross-sectional area and P the perimeter. For the nasal cavity, both P and A are summations of both the left and the right side of the cavity. $U_{avg}$ is the area averaged velocity at the cross section. The result is plotted in Fig 7.

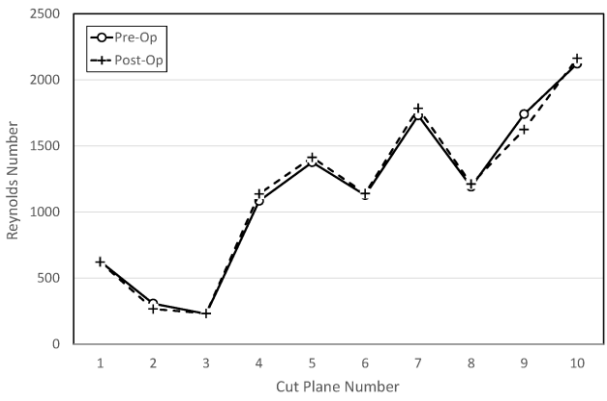

**Figure 7**: Calculated Reynolds numbers at selected cut planes (see Fig. 8) based on area averaged velocity and hydraulic diameter of the cut plane

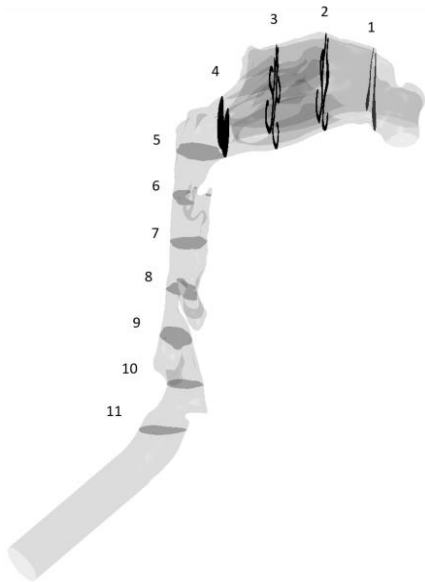

**Figure 8**: Location and numbering of cut planes used for calculation of Reynolds number.

The Reynolds number ranges from 621 to 2160, which is within the laminar regime. Based on this, the laminar approach is suitable.

**Pressure**

At a flow rate of 250 ml/s, the calculated pressure drop from inlet to the larynx is 34.46 Pa pre-operatively, and 44.56 Pa post-operatively. This means that a greater pressure difference and more effort are needed to inhale the same amount of air after surgery. The major change in pressure drop is found over the nasal cavity. This has increased with 5.41 Pa after surgery. The pressure on the wall in the nasal cavities can be seen in Fig. 12.

The major change in the pressure distribution after surgery is the high pressure gradient at the smallest cross section in the anterior nose. This change can be seen on both sides post-operatively, and is the main reason for the total change in pressure drop over the nasal cavities after surgery. Besides from this pressure change in the anterior nose, the pressure development follows the same trend pre- and post-operatively, but the pressure is overall lower post-operatively. Another change between the two models can be observed at the posterior laryngopharynx. The pressure drop at this region was lower before surgery.

**Nasal Resistance**

The nasal resistance was measured with RRM and RMM both pre- and post-operative. Pre-operative, measured results are only available for the right nasal cavity. This is because the nasal passage was too narrow for the tests to work. Post-operative measurements are available for both sides. However, even post-operatively the results are limited and only available for measurements after decongestion. Both measured and CFD results are presented in Table 1.

**Table 1:** Nasal resistance, measured and calculated results.

|  | Flow rate [ml/s] | R, RRM [Pa s/ml] | R, CFD [Pa s/ml] |
|---|---|---|---|
| Pre, right | 143.8 | 0.1732 | 0.0429 |
| Pre, left | 106.2 | Not Measured | 0.0581 |
| Post, right | 135.1 | 0.1145 | 0.0888 |
| Post, left | 114.9 | 0.6167 | 0.1044 |

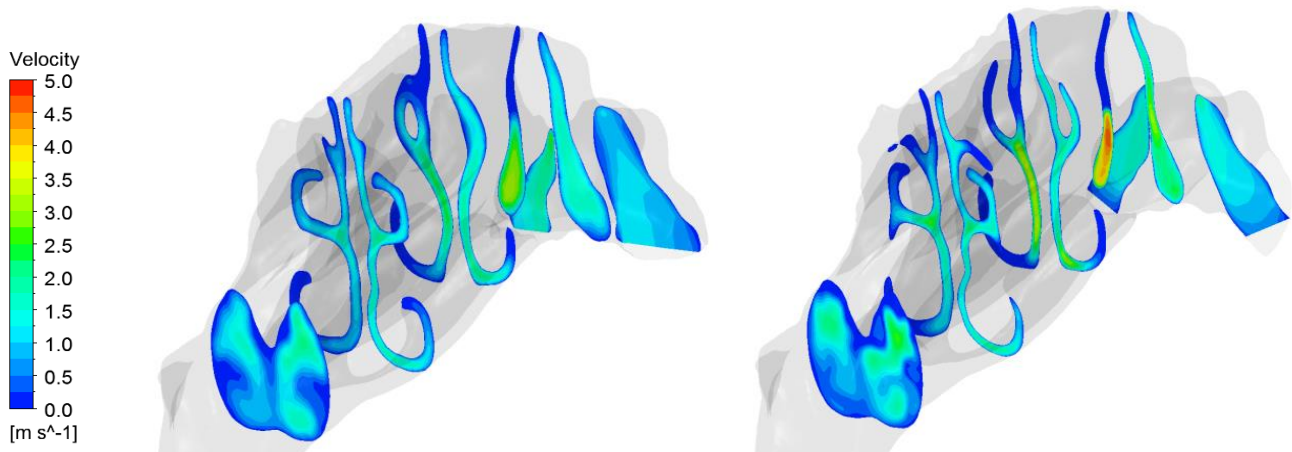

**Figure 9:** Contour plot of the velocity across coronal cross sections in the nasal cavity pre-operative (left) and post-operative (right). The models are viewed from the right side.

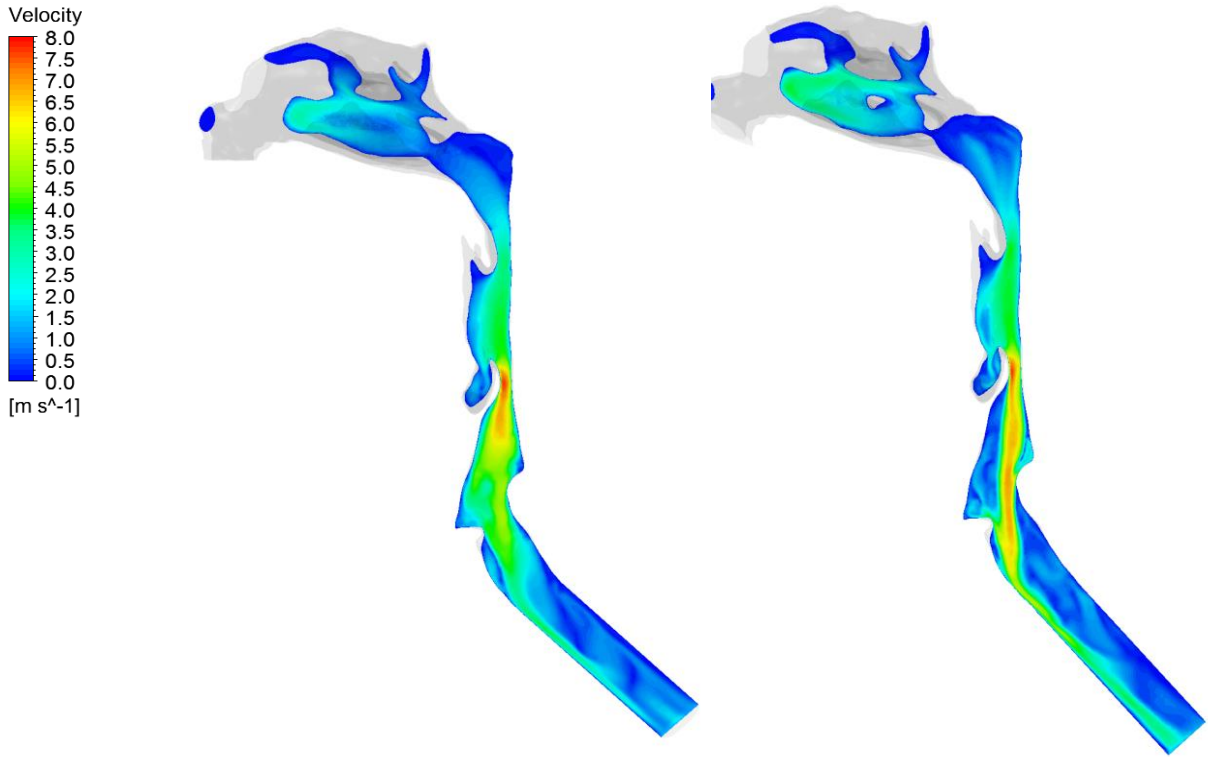

**Figure 10**: Contour plot of the velocity of the velocity across a sagittal cut plane at the middle of pharynx and larynx, and the left nasal cavity. The pre-operative results to the left, and the post-operative results to the right.

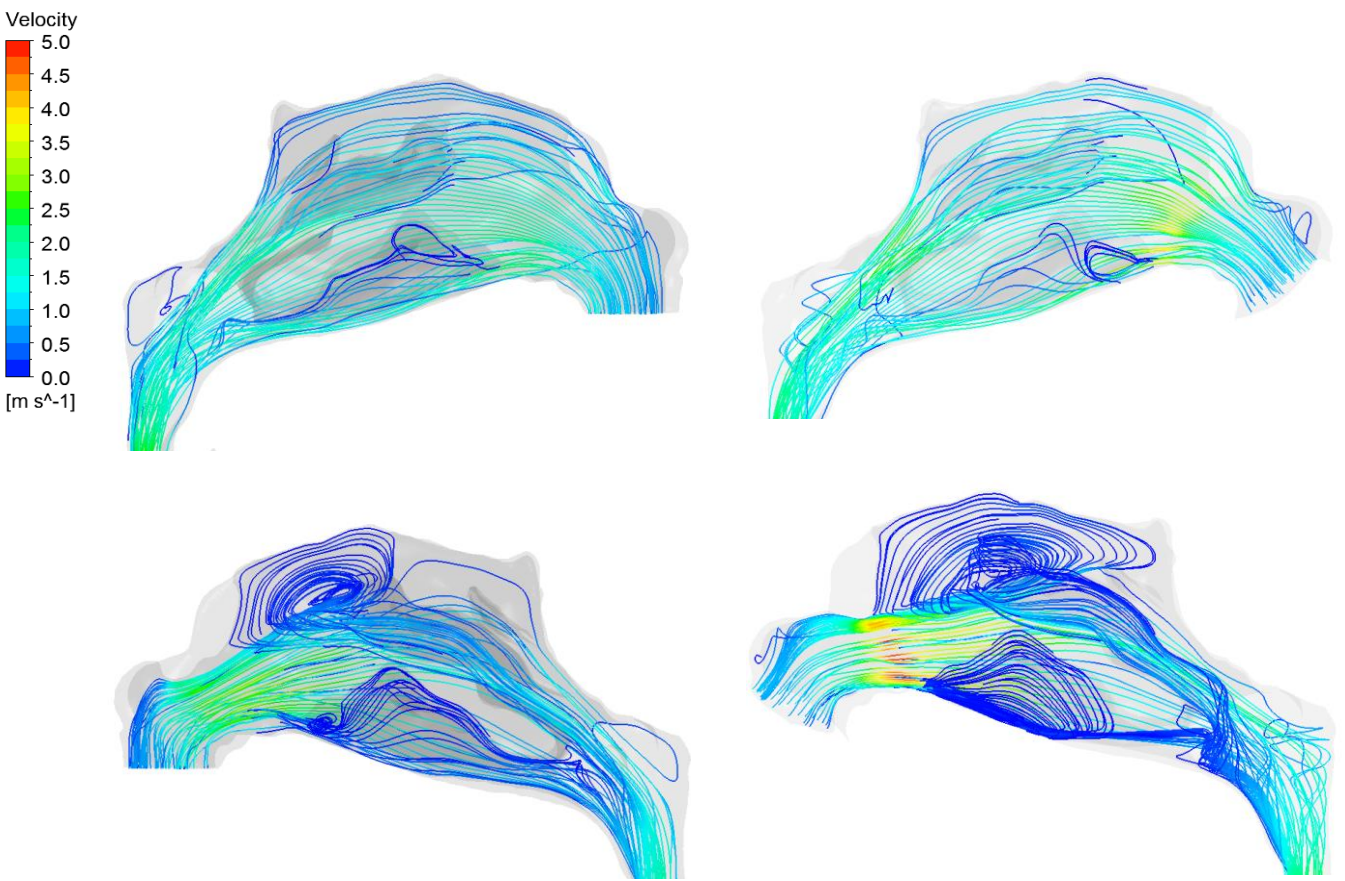

**Figure 11**: Velocity streamlines in the right (upper) and left (lower) nasal cavity pre (left)- and post (right)-operative.

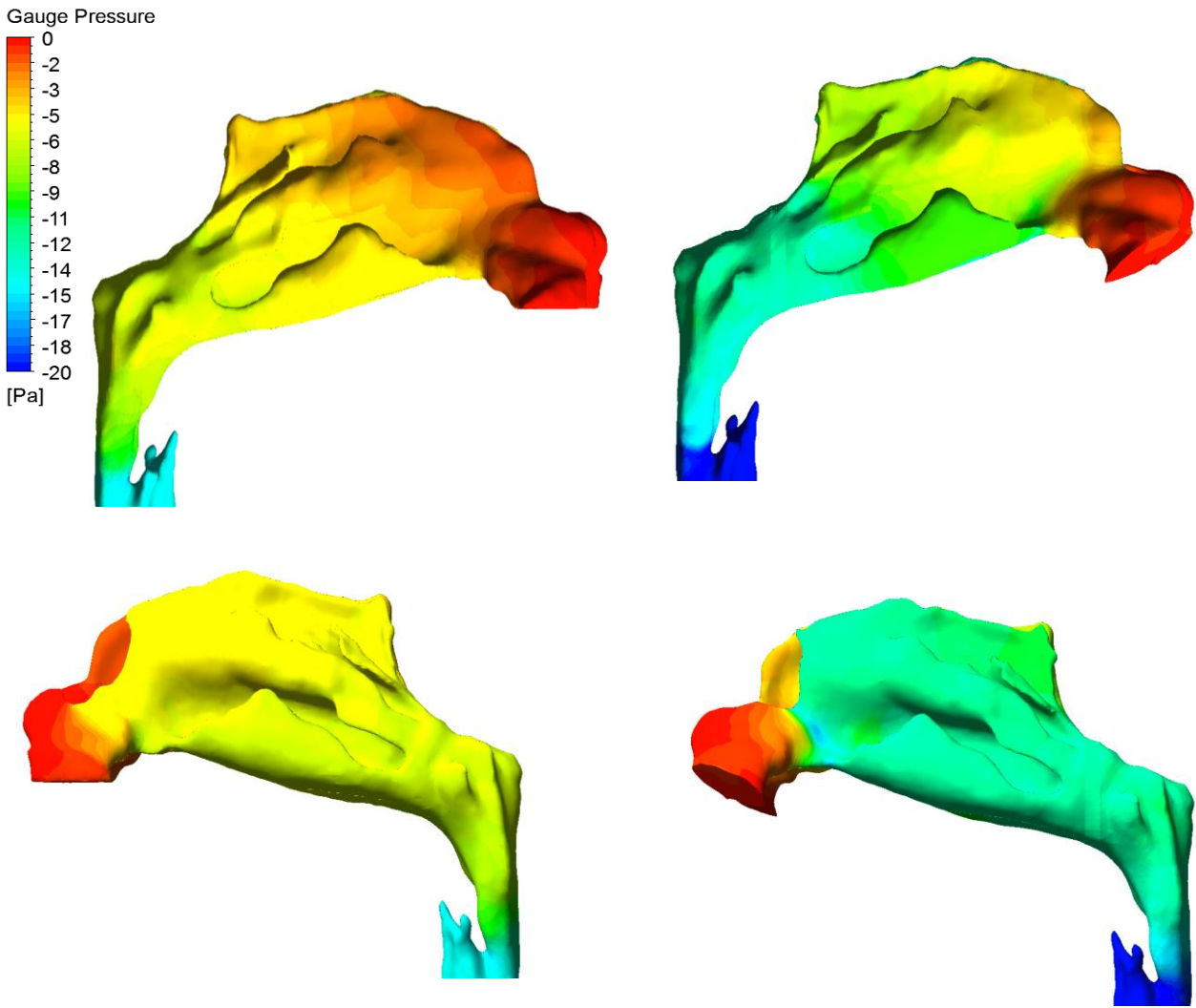

**Figure 12**: Contour plots of the pressure distribution at the wall in the nasal cavity pre (left) – and post (right)-operative. The nasal cavity is viewed from the right (upper) and left (lower).

## DISCUSSION

### CFD and geometry approach

The CFD results show airflow-patterns similar to previous modelling work and experiments reported in the literature. However, the calculated nasal resistance values from the CFD results differ remarkably from the rhinometric measurements. A possible source of error is the CFD approach. This study established a laminar base-case, and based on the Reynolds number, the flow is within the laminar flow regime. However, the geometry of the upper airways varies greatly and turbulent effects may be present at certain regions. The turbulent approach has been studied by Aasgrav (2016) and Aasgrav et.al (2017). The CFD-simulations with laminar and turbulent models gave similar results for both pressure and velocities and indicate that the errors must be related to earlier steps in the modelling procedure such as geometry retrieval and/or CT acquisition.

As pointed out (Shreck et al., 1993) the nasal resistance is highly dependent on the cross sectional area. A comparison of the measured and calculated hydraulic diameter on the pre-operative model shows that the hydraulic diameter is significantly larger (about 60%) in the CFD model (decongested RMM). This may account for the large deviation between the measured and calculated nasal resistance. A decrease in hydraulic diameter by 60% is approximately a layer of one voxel off the model (0.3mm on the left side, and 0.5mm on the right side). When reducing the geometry, by a trial and error approach, the reduction of a voxel layer corresponds to an upper HU-value of approximately -600 HU. It should be noted that this value is higher than those reported in the literature, but nevertheless, this strongly indicates that the geometry is too large, and that the HU-values for segmentation should be reevaluated. Finding a suitable HU-range by calculating the hydraulic diameter after simulating the air flow is easy, but not ideal. Predicting the right HU-range and segmentation procedure earlier on in the process is challenging, and a standard approach for setting the upper HU-value is needed. However, more work is needed in order to study the effect of a reduction of the hydraulic diameter, and the sensitivity to HU values. Overall, it should be noted that the segmentation procedure is both time consuming and prone to human errors as a large amount of manual segmentation is needed.

As mentioned, the model was smoothed in order to reduce digitalization artefacts. This creates a smooth surface, with less friction on the walls. By doing so, an idealized, and perhaps unrealistic, nasal cavity may be created. In reality, the walls of the nasal cavity are covered by mucosa and nasal hair in the anterior nose, which will make a more irregular surface. While eliminating the unrealistic stair-steps of the model created by digitalization, the smoothing may have resulted in much less friction than in the real case. This may have contributed to the overall low nasal resistance, but does not explain the increase in pressure drop over the nasal cavity after surgery.

Another simplification of the model is that all the walls are assumed to be rigid. For most of the nasal cavity this is a good approximation, but the pharyngeal walls are known to be less rigid. By modelling these walls with fluid-structure-interaction (FSI), effects not captured by the CFD approach so far may be evident. In particular, a collapse of the airways during inspirational flow may occur. However, the hysteresis effects in RMM data are expected to be negligible as the calculated flow resistance behaves similar during inhalation and expiration.

**Physiological effects**

Whether or not the difference in measured and calculated hydraulic diameter is solely based on the segmentation procedure, or if there actually is a physical difference between the patient at the time of CT scans and at the time when the rhinometric tests were taken is not yet clear. In addition to the segmentation procedure, the geometry difference may also be caused by physical effects captured on CT. One concern is the nasal cycle – of which the effect it has on the model is not yet determined. CT gives an instantaneous representation of the upper airway, but the geometry of the upper airway is in fact constantly changing because of the nasal cycle. The cyclic movement works in a way so that the volume of the left and the right side most of the time is asymmetric. This means that CT from the same day can give different geometries. When comparing pre- and post-operative CT data, the differences in the geometry can be greater or smaller depending on which nasal cavity is dominant at the time. A comparison of the CT images shows indications of the patient being in different cycles pre- and post-operatively. A comparison of a coronal slice in the anterior nasal cavity can be seen in Fig. 13.

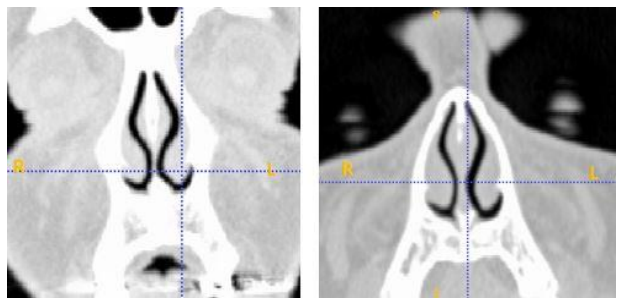

**Figure 13:** Coronal CT view of the nasal cavity pre (left) - and post (right) -operative showing indications of a change in the nasal cycle in pre- and post-operative CT.

The right side appears to be larger in the pre-operative CT than in the post-operative one, and hence there is a more distinct difference between in the left and right nasal cavities after surgery. This also corresponds with the hydraulic diameter that is especially large on the right side. The need for taking the nasal cycle into account when modelling the nose has been pointed out in previous studies. Patel et al. (2015) compared pre- and post-operative models to study nasal airway obstruction and had to limit their study subjects to those that seemed to be in the mid cycle (symmetric) of the nasal cycle both pre- and post-operatively. In order to include more subjects into the study, they came up with a method for modelling the nasal cycle. By changing the thickness of the inferior and middle turbinate in addition to the septal swell body, the nasal cycle is taken into account. Patel found that the surgical effect was more correctly simulated when the geometry has been adjusted to eliminate the influence of the nasal cycle. Another option is to simply try to avoid the nasal cycle as a source of error. This could be done by obtaining the CT after the patients nose have been decongested by nasal spray. It is, however, important to keep in mind that the decongested state is unnatural. When measuring the AHI during a sleep study, the nasal cycle is present, and including the nasal cycle in the model instead of eliminating it may give a more realistic result. As long as the nasal cycle is ignored, the CT scan image data can make show different geometries of the same nasal cavity, and make it more difficult to reproduce data.

The high reduction in AHI measured clinically is not as clearly observed in the CFD results. The simulation results show a significant change in the flow patterns in the nasal cavities, but only a small change in the flow patterns in the pharynx and larynx between the pre- and post-operative models. The major differences after surgery are a more evenly distributed flow between the two nasal cavities, and an increase in the pressure drop over the nasal cavity. The change from mouth breathing to nasal breathing can be the cause of the major improvement in AHI. It can be hypothesized that the obstructions in the nose of the patient made it too difficult to breathe through the nose, and that he instead was breathing through his mouth during sleep. When breathing through the mouth (and opening the mouth), the volume in the pharynx decreases as the tongue and soft palate moves posterior towards the pharyngeal walls. This might even close the pharynx, and can result in both apneas and hypopneas which can explain the high AHI reported before surgery. After surgery, the simulation results show a more symmetrical flow in the nose which might make it easier to breathe through the nose - perhaps enough for the patient to breathe only, or mostly, through his nose. If this is the case, the pharyngeal volume will be significantly larger than it is when the mouth is open, and the risk for collapse will be reduced. It is, however, not known if the patient changed from mouth- to nasal breathing after surgery as there are no available data for this, and more information about the patients sleeping habits is needed to verify this. If the patient did sleep with an open mouth, the geometry should also include the oral cavity (a third inlet), and a CFD study on that geometry should be included as well to relate the CFD-results with the AHI. Modelling of open mouth breathing calls for a more complex model as this requires the soft palate to be movable, and FSI is needed in order to do so.

It should be noted that all the mechanisms of OSAS are not known and understood, and there can be other mechanisms causing apneas and hypopneas that are not visible by studying the flow. It has been suggested that neurological mechanisms also may influence the breathing pattern. It is currently unknown how this can be affected by nasal surgery (Moxness and Nordgård, 2016). A last remark is that the CT data, which is the basis for the numerical simulations, is obtained when the patient is awake, while the AHI is measured during a sleep study. During sleep, the muscles relax and the pharyngeal wall can become narrower as the muscles that are supporting it are relaxed. In addition, the muscle-relaxation may also make the tongue relax and fall posterior, when sleeping in the supine position, due to gravity.

## CONCLUSIONS AND FURTHER WORK

In the current paper, the airflow in the human upper airways has been simulated for an OSAS patient to study the effect of intranasal surgery for alleviation of OSAS. A base-case with laminar flow was made and the results from CFD were compared with clinical measurements, in particular measurements of the nasal resistance. The CFD results and the measured results did not correspond, and the main errors are expected to be caused by differences between the geometry of the upper airway and the airway being modelled. Further work will be focused on making an anatomical correct geometry before proceeding further with numerical simulations.

## ACKNOWLEDGEMENTS

This project work is part of a collaborative research project, "Modeling of Obstructive Sleep Apnea by Fluid-Structure Interaction in the Upper Airways", between NTNU, SINTEF Materials and Chemistry, and St. Olavs Hospital, Trondheim University Hospital, Norway. The project is funded by the Research Council of Norway, under the FRINATEK program (OSAS, 2017).